\documentclass[12pt]{article}
\usepackage[latin1]{inputenc}
\usepackage{graphicx}
\usepackage{epsfig}
\begin{document}

\title{Dipole states in stable and unstable nuclei 
}

\author{D. Sarchi\footnote{Present address:
Institute of Theoretical Physics, CH 1015 Lausanne-EPFL, Switzerland.}, 
P.F. Bortignon and G. Col\`o\\
Dipartimento di Fisica, Universit\`a degli Studi\\
and INFN, Sezione di Milano, Via Celoria 16, 20133 Milano, Italy}

\date{\today}

\maketitle

\begin{abstract}

A nuclear structure model based on linear response theory (i.e., Random
Phase Approximation) and which includes pairing correlations and
anharmonicities (coupling with collective vibrations), has
been implemented in such a way that it can
be applied on the same footing
to magic as well as open-shell nuclei. 
As applications, we have chosen to study the dipole
excitations both in well-known, stable isotopes like $^{208}$Pb and $^{120}$Sn as
well as in the neutron-rich, unstable $^{132}$Sn nucleus, by 
addressing in the latter
case the question about the nature of the low-lying strength.
Our results suggest that the model is reliable and predicts in all cases
low-lying strength of non collective nature. 

\end{abstract}

\leftline{PACS numbers: 24.30.Cz, 21.60.Jz, 27.60.+j}

\vspace{1.5cm}

Giant Resonances (GR) in atomic nuclei lie at excitation energies
above the nucleon separation threshold ($\sim$8-10 MeV), have different
multipolarity and carry different spin-isospin quantum numbers. They are
associated with the elastic, short-time 
($\hbar /10 $ MeV $\approx 10^{-22}$
s) response of the nuclear system. They have been observed throughout the
mass table with large cross sections, close to the maximum allowed by sum
rule arguments, implying that a large number of nucleons participate in a
very collective nuclear motion \cite{BBB,HVW}. 

The giant dipole resonance (GDR) has been the first to be discovered, and is probably
the most studied among the various nuclear collective vibrations. In well-bound
systems it is often depicted as a coherent oscillations of essentially all protons against
all neutrons, boosted by an initial displacement of the centers of the two
distributions induced by an external electromagnetic field. 
The energy of this resonance in MeV, 
is given by the empirical 80 A$^{-1/3}$
law.

In the 
neutron-rich isotopes, it has been suggested that part of the energy provided by
the external field can be used to excite a different kind of motion. In fact, if
the valence neutrons occupy orbitals with lower energies and larger radii than the
inner, or "core", particles, these orbitals must be somehow decoupled from the
others. Consequently, the excitations of the valence neutrons should result in
vibrations of these particles against the inner ones, namely in modes with
different characteristics than the standard GDR. 

Recently \cite{GSI} a systematic study in the neutron-rich isotopes
$^{18-22}$O, via relativistic Coulomb excitation, has identified sizeable
strength (about 10\% of the Thomas-Reiche-Kuhn (TRK) sum rule) below 15 MeV, that is, well 
below the GDR region. The question about the collectivity of this strength
is left to theorists. So far, every microscopic model (either within
the non-relativistic \cite{NP01}  or relativistic \cite{Vret}  
framework) seems to find only non-collective states in the low-lying
region.  
A further natural question which arises, is whether this 
low-lying dipole
strength
systematically exists in the medium-heavy nuclei and whether 
a certain degree of collectivity can develop, with increasing
mass number. 
Another issue is whether the properties of the low-lying dipole
can shed light on the (possible) modifications of the nuclear
effective Hamiltonian(s) when one goes far from the line of stability. 

These questions have been attacked within the most widely used mean field
models \cite{ProcHanoi}, which however do not take into account the 
spreading effects, as we will do in this work.

In fact, in nuclei 
the mean field defines a surface which can vibrate leading to a
spectrum of collective low-lying excitations (phonons). The coupling of the
nucleons to these dynamical vibrations, a process which goes beyond
mean-field, can strongly renormalize the single-particle motion by changing
the energy and the occupancy of the levels around the Fermi energy and,
eventually, by providing a single-particle spreading width 
$\Gamma^\downarrow_{s.p.}$. In keeping with the
fact that the giant resonances 
can be viewed as correlated particle-hole (p-h) or two quasi-particle 
(2qp) states, the very same
mechanism will produce their spreading width $\Gamma^\downarrow_{GR}$, 
although
quantitatively this width is reduced compared to the sum of the 
particle and hole widths since this is the strong effect of 
the coherence between the particle and hole motions as shown in 
Ref. \cite{revmodphys}. 

The aim of this paper is the following. 
If on one hand mean field studies are nowadays performed using
sophisticated forces well linked to nuclear matter properties, 
they do not include the phonon
coupling mentioned above.
We would like to provide a unified description of the properties
of the giant resonances within a self-consistent model which
includes such a coupling in both magic and open-shell nuclei. 
In this sense, the work is a continuation of the
old pioneering calculations of Ref. \cite{NP81}, with the 
very clear improvement coming from the self-consistent use of
an effective nucleon-nucleon interaction. 
Among other limitations, in Ref. \cite{NP81} pairing was considered in a 
simplistic way and it was not possibile
to calculate the giant resonances along an isotopic chain. As far as 
the particle-hole channel is concerned, a consistent calculation of
the dipole states was 
performed in Ref. \cite{NP01}. In this work, a Skyrme-type interaction 
was employed and the residual interaction between quasiparticles
was self-consistently derived. However, the pairing contribution 
was dropped from the residual interaction. 
Nowadays, since extrapolations to neutron-rich nuclei are needed,
it is desirable to be able to reproduce the lineshape of the
giant resonances in different nuclei without {\em ad hoc}
adjustments, neither in the particle-hole nor in the pairing
channel.
Thus, we have accomplished the implementation a more general model 
compared to that of \cite{NP01}, in the sense that we now treat
consistently also the pairing interaction. 
We test this new model, that we call QRPA-PC (Quasi-particle Random 
Phase Approximation plus Phonon Coupling), in the case of dipole.
We analyze the well-known $^{208}$Pb nucleus already, 
and also very recently \cite{Rich}, investigated 
in many aspects, and also 
another stable, yet open shell nucleus, namely $^{120}$Sn. In both
cases we can compare with measured photoabsorbtion cross sections. 
Then, we make predictions for the 
neutron-rich isotope $^{132}$Sn, for which experimental results should
be soon available.  

Our starting point is a mean field calculation performed in the 
HF-BCS (Hartree-Fock-Baarden-Cooper-Scrieffer) plus QRPA approach.
As already mentioned, 
the model is self-consistent, in the sense that 
a Skyrme force is used in the 
particle-hole channel, while in the particle-particle 
one a zero-range pairing interaction 
\begin{equation}
V_{pair}(\vec{x}_1,\vec{x}_2)=
-V_{0}\left[1-\left(\frac{\rho(\frac{\vec{x}_{1}+\vec{x}_{2}}{2})}
{\rho_{c}}\right)^{\gamma}\right]\delta(\vec{x}_{1}-\vec{x}_{2})
\label{eq:pairint}
\end{equation}
is adopted. The parameters of this pairing force are adjusted
in such a way that the average pairing gap resulting from the HF-BCS calculation
(where the average is made by considering levels within $\pm$ 5 MeV from the
Fermi energy) agree with  
the result of the three-point formula 
(the so-called $\Delta^{(3)}_{odd}(N+1)$ formula \cite{delta3}). 
The properties of
the excited states are not considered as input to fit the force parameters.
For the tin isotopes, using in the p-h channel the 
SIII force \cite{Beiner}, we
obtain that the agreement of the average paring gap with the outcome of 
the three-point formula is within $\sim$100 keV in $^{120-126}$Sn 
(if $\gamma=1$) with $\rho_{c}=0.16$ 
fm$^{-3}$ and $V_{0}=640$ MeV fm$^{3}$. 
In fact, pairing is only included in the case of $^{120}$Sn. For $^{132}$Sn
and $^{208}$Pb, HF-BCS plus QRPA is replaced by HF plus RPA. 

The HF-BCS equations are solved in coordinate space. The step and
upper limit for the radial coordinate are equal respectively to 0.1 fm and
25 fm. The continuum part of the
single-particle spectrum has been discretized 
by using box boundary conditions. Only the neutron single-particle states 
1h$_{11/2}$, 2f$_{7/2}$, 1h$_{9/2}$, 1i$_{13/2}$, 
3p$_{3/2}$ are considered for the BCS equations. 
The RPA or QRPA equations are solved in the configuration space using
the same technique as explained in \cite{NP01}.  
We have 
found that the results are stable and the energy-weighted sum rule (EWSR) 
is 
satisfied if an upper cutoff in the two quasi-particle energies
is set at 100 MeV. 

In this self-consistent mean field model,
the spurious states $0^{+}$ and $1^{-}$ associated 
respectively with the breaking in the
superfluid ground state of the
number of particles, and with the translational symmetry, 
appear at zero energy
when the p-h matrix elements are renormalized by a few percent
(3\%), because of the finite basis size. 
This little renormalization does not 
affect significantly 
the dipole response at any energy. No renormalization of the p-p interaction
is needed for the $0^+$ state.

To include the coupling with the vibrations, we use the same Hamiltonian
of Ref. \cite{GL01} in the full neutron-proton scheme.
The formalism has been described in detail in Ref. \cite{NP01}, to which
we confer the reader (cf. in particular Sec. 2 and Appendix A). 
In the present case the phonons of our model are the states resulting from
RPA or QRPA calculations of multipolarity 1$^{-}$, 2$^{+}$, 3$^{-}$ and 
4$^{+}$ having energy below 30 MeV and exhausting at least 5\% of isoscalar 
or isovector strength. The properties of the low-lying 
2$^{+}$ and 3$^{-}$ states in the Sn
isotopes are discussed in ref. \cite{NP03}.    

We have checked by means of an explicit 
calculation for the $^{208}$Pb case, that 
continuum-RPA gives peak positions quite
similar to our discrete RPA, with escape widths smaller than 0.5 MeV, 
in agreement with the experimental data \cite{HVW}.

We briefly discuss 
the sensitivity of the dipole results to the particular Skyrme 
force employed, before analyzing in detail the results obtained
using the parametrization SIII.
The issue has been studied in Ref. \cite{GLPL}. In that work, it 
was found by exploting 
the $^{208}$Pb case that the 
RPA centroid energy $E_{-1}$ (defined as $(m_1/m_{-1})^{1/2}$,
where $m_k$ is the $k$-th moment of the strength function) 
of the dipole, has a linear dependence on the quantity
$F_D$ associated with the parameters of the Skyrme force in question, 
\begin{equation}
E_{-1}=\lambda + \mu F_D.
\label{linearE1}
\end{equation} 
The explicit expression of $F_D$ can be found in Ref. \cite{GLPL}, but we
remind here that it contains essentially the square root of the 
product of 
the bulk symmetry energy
$a_\tau$ (sometimes indicated by $J$ or $a_4$), of the 
ratio between the surface symmetry
energy coefficient $a_{ss}$ and $a_\tau$, and 
of the effective mass $m^*$. The dimensions of $F_D$ are MeV$^{1/2}$.  

It is confortable to have found in the present context
that the linear relation (\ref{linearE1}) is valid also for 
$^{120}$Sn. While in the $^{208}$Pb case, the values $\lambda$=11.35 MeV
and $\mu$=0.77 MeV$^{1/2}$ were obtained in \cite{GLPL}, 
in the case of $^{120}$Sn 
($^{132}$Sn) we find in the present analysis $\lambda$=9.28 MeV 
(11.92 MeV) and $\mu$=2.34 MeV$^{1/2}$ (1.46 MeV$^{1/2}$).
For $^{120}$Sn, forces like SkP \cite{Doba}, SLy4 \cite{Chab} and 
SGII \cite{SGII} give results 
for the peak energy which are lower than the experimental ones. This is
not the case for the force SIII, and we will use it, since the phonon coupling 
induces a downward shift of the RPA peak energy.

To be more quantitative, in Table \ref{tab:unpRPA},
we compare the mean field results obtained with the 
SIII and SLy4 forces. 
We report the centroid energies 
$E_0=m_{1}/m_{0}$
obtained within RPA or QRPA, together with the unperturbed HF or HF-BCS 
results, 
and with the phenomenological predictions.  
We can observe that, for all nuclei we consider, the unperturbed 
value of the 
centroid obtained using SIII is similar 
(yet slightly lower) than that obtained using SLy4,
whereas the RPA result is sistematically higher.
In fact, while for the two forces the effective 
mass is about the same ($m^{*}/m=0.76$ for SIII and 
$m^{*}/m=0.7$ for SLy4), 
the repulsive  
matrix elements obtained with SIII are sistematically larger  
that those of the SLy4.
From the comparison between RPA and QRPA
results for the nucleus $^{120}$Sn, we can deduce that the 
pairing interaction does not play a crucial role, as expected. 

\begin{table}
\begin{center}
\begin{tabular} {|c|c|c|c|}
\hline
nucleus & SIII & SLy4 & 80 A$^{-1/3}$ (41 A$^{-1/3}$) \\
\hline 
$^{208}$Pb & 14.9 (9.4) & 13.4 (9.8) & 13.5 (6.9) \\
\hline
$^{120}$Sn (RPA) & 17.1 (10.9) & 15.0 (11.0) & 16.2 (8.3) \\
\hline
$^{120}$Sn (QRPA) & 17.3 (11.3) & 15.1 (11.6) & 16.2 (8.3) \\
\hline
$^{132}$Sn & 17.0 (10.7) & 15.2 (11.0) & 15.7 (8.1) \\
\hline
\end{tabular}
\end{center}
\caption{Centroid energies $E_0=\frac{m_{1}}{m_{0}}$ obtained within 
RPA or QRPA (together with the unperturbed HF or HF-BCS result in 
parenthesis), compared with the empirical prediction 80 A$^{-1/3}$ 
(41 A$^{-1/3}$).}
\label{tab:unpRPA}
\end{table} 

In Table 2, the results for the centroid energy without and with the
phonon coupling are shown.
In the case of $^{208}$Pb, the centroid energy after the 
coupling is 14.4 MeV, to be compared with the value 
14.9 MeV at the
level of RPA. 
In the case of $^{120}$Sn, the centroid energy, 
at the level of the complete calculation 
is the same value of the QRPA calculation. 
The same is true for $^{132}$Sn. 

\begin{figure}[htbp]
\begin{center}
\includegraphics[scale=0.5,angle=-90]{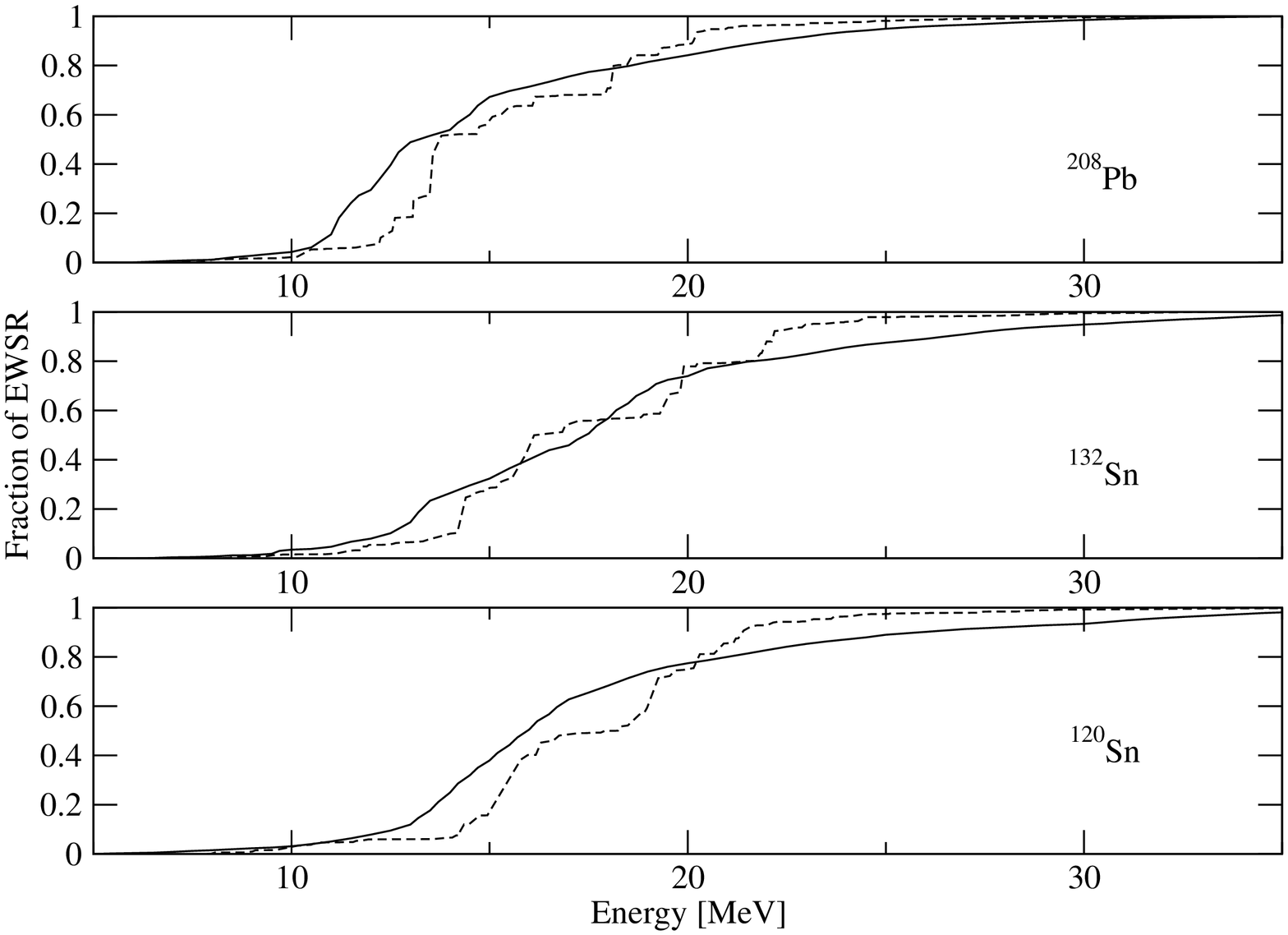}
\end{center}
\caption{Exhaustion of the dipole EWSR in the three nuclei considered in the 
present paper. The full line refers to the complete (Q)RPA 
calculation, while the
dashed line to the (Q)RPA calculation.}
\label{fig0:ewsr}
\end{figure}

The integral of the strength as a function of the upper integration limit, 
that is, the cumulated value of the EWSR, is shown in Fig. \ref{fig0:ewsr}.
This cumulated value is shown as a fraction of the total expected 
value, which is the TRK sum rule multiplied by the enhancement factor
$(1+\kappa)$. The values of the total EWSR in MeV$\cdot$fm$^2$ (and 
of $\kappa$ in parenthesis) are, for $^{120}$Sn, $^{132}$Sn and 
$^{208}$Pb respectively: 2448 (0.399), 2566 (0.397) and 4202 (0.408). 
One can see that, in order to exhaust more than 90-95\% of the  
EWSR in the (Q)RPA-PC calculation, one has to reach 30 MeV in the
case of $^{208}$Pb and 40 MeV in the case of the Sn isotopes. 
We stress that, while the shape of the strength distribution 
and the GDR peak position
change when the particle-vibration coupling is taken in account,
the centroid and the EWSR values are much less affected.

In the Figures 2-4, we show the calculated photoabsorption cross sections,
together with the experimental results for the two stable nuclei.
The values of energies and the widths obtained from a lorentzian fit 
of our cross sections are summarized in Table \ref{tab:width}.

We can see that the agreement with the experimental results is good for
both $^{208}$Pb and $^{120}$Sn. 
This gives us confidence to use the model for a prediction of the 
photoabsorbtion in the unstable nucleus $^{132}$Sn.
In this case the width is about 6 MeV, like in $^{120}$Sn, but 
the cross section is very fragmented. 

In the low-lying energy region, we obtain strength
in the three nuclei, as we discuss below.
The cumulated value of the EWSR below 12 MeV is shown in 
Fig.\ \ref{fig:ewsr2}.
For $^{208}$Pb, a comparison is possible with the data of Ref. \cite{Rich}.
Our calculation gives a total integrated strength B(E1)=0.53 e$^{2}$fm$^{2}$
(below 6 MeV) and B(E1)=1.27 e$^{2}$fm$^{2}$ (up to 8 MeV),
whereas the experimental data are B(E1)=0.52 e$^{2}$fm$^{2}$
(below 6 MeV) and B(E1)=0.80 e$^{2}$fm$^{2}$ (up to 8 MeV).
The states carrying this strength are mainly single-particle excitations,
so the value of their energy is affected by
the single-particle energy spectrum.
On the other hand, the global B(E1) distribution is well reproduced
qualitatively and quantitatively.

Coming to the Sn isotopes, the low-lying dipole strength grows 
faster, at the level of (Q)RPA, in $^{120}$Sn. The phonon 
coupling affects more $^{132}$Sn, reversing somewhat the trend.  
Also from Figs. \ref{fig3:132} and \ref{fig2:120}
one can see that in $^{132}$Sn four definite peaks exhaust the strength
below 10 MeV, whereas in $^{120}$Sn only a kind of smooth background is
visible. 

We have analyzed the structure of the four lying below 10 MeV in
$^{132}$Sn. In RPA, the lowest peak at 8.44 MeV is mainly (78\%) associated 
with the 3s$_{1/2}$ $\rightarrow$ 3p$_{3/2}$ single-particle transition,
and absorbs only 0.2\% of the EWSR. There are two higher peaks
which lie at 8.61 MeV and 9.53 MeV and absorb respectively
0.5\% and 0.3\% of the EWSR: they involve an admixture of the 
2d$_{3/2}$ $\rightarrow$ 3p$_{1/2}$ and 3s$_{1/2}$ $\rightarrow$
3p$_{1/2}$ transitions. Other small states involve the d $\rightarrow$ f
transitions but there is no peak to which more than two or three components
contribute. The pattern is very similar if we move to the 
interaction SLy4. This situation is somewhat at variance with the 
findings of Ref.\ \cite{Vret}, in which the lowest state emerging
from the relativistic RPA has four or five configuration well
mixed and it exhausts alone about 1.4\% of the EWSR. We conclude that
an experiment aimed to explore the existence of either a well defined, 
relatively collective peak or a fragmented pattern in the low-lying
energy region can really discriminate among microscopic self-consistent
models.

\begin{table}
\begin{center}     
\begin{tabular} {|c|c|c|}
\hline
nucleus & RPA & QRPA-PC \\
\hline 
$^{208}$Pb & 14.9 & 14.4 \\
\hline
$^{120}$Sn & 17.1 & 17.1 \\
\hline
$^{132}$Sn & 17.0 & 17.0 \\
\hline
\end{tabular}
\end{center}
\caption{Centroid energies $E_0=\frac{m_{1}}{m_{0}}$ in MeV, obtained in 
the (Q)RPA and (Q)RPA-PC calculations. Only in the case of $^{208}$Pb we 
observe an appreciable shift when the phonon coupling is introduced. In the
other two cases the downward shift is smaller than 100 keV.}
\label{tab:centroid}
\end{table} 

\begin{table}
\begin{center}
\begin{tabular} {|c|c|c|c|}
\hline
nucleus & QRPA-PC & exp. \\
\hline 
$^{208}$Pb & 13.1 (3.7) & 13.46 (3.9) \\
\hline
$^{120}$Sn & 15.7 (5.3) & 15.4 (4.9) \\
\hline
$^{132}$Sn & 15.5 (5.8) & - \\
\hline
\end{tabular}
\end{center}
\caption{Values of the peak energy (width) in MeV 
calculated in (Q)RPA-PC (calculated by means of a lorentzian
fit to the cross sections) in comparison with the experimental 
values.}
\label{tab:width}
\end{table}

 
The analysis of the RPA-PC wave functions of the states is slightly more
difficult and less transparent, also because of the use of a complex 
Hamiltonian. However, we can state that the wave functions do not
acquire more collectivity. This is confirmed by considering the
transition densities, which are reported in Fig.~\ref{fig4:td} and
are associated indeed to the RPA-PC calculation. The transition density 
associated with the the GDR peak at 13.7 MeV is compared with that of 
the low-lying state at 9.7 MeV. This latter has many nodes while that 
of the GDR displays the expected shape of collective type. The low-lying 
transition densities is also dominated by the neutron contribution 
at the surface and this fact is of course relevant for any consideration 
concerning its excitation by means direct reactions. The neutron
character of the low-lying state at the surface is even more
pronounced in the complete calculation, than in simple RPA. 

To judge the collectivity of the low-lying strength, one may 
think about using the so-called ``cluster sum rule'' $S_{clus}$\ \cite{CSR}. 
However, as stated in many occasions, its use is jusitified only
provided an unambigous decoupling of excitations of core and valence
particles is present in the physical systems. Already from the 
above discussion, and in particular from the analysis of the wave functions,
it is clear that this is not the case for $^{132}$Sn: all the excess
neutrons contribute to the low-lying strength and not only the least
bound ones. 
 
In conclusion, we have implemented with the present work a complete
model which includes states of four quasi-particles nature in the 
description of the vibrational nuclear states. The model makes
self-consistent use of a Skyrme force in the particle-hole channel and
of a zero-range, density-dependent pairing force in the particle-particle
channel. We have applied this model to the description of the the
dipole strength in different systems, reaching satisfactory results.
In $^{120}$Sn and $^{208}$Pb we obtained agreement with the experimental
data, for the peak energy and width of the GDR (and for the pygmy states
recently studied in $^{208}$Pb). In $^{132}$Sn, we have predictions 
which show a large fragmentation of the main GDR and the absence of
collective states in the low-lying region. This latter result, 
obtained using two different effective forces, is somehow at variance
with the outcome of relativistic RPA studies\ \cite{Vret}. However, it 
is consistent with the discussion in Ref.\ \cite{HH}, 
where it is argued that the soft dipole strength,
observed in halo light nuclei, should decrease in skin nuclei,
due to the coupling to the GDR. These remarks point to the importance
of clear-cut experimental data which are able to identify or exclude
the presence of a low-lying collective dipole in medum-heavy nuclei.

\newpage

\begin{figure}[htbp]
\begin{center}
\includegraphics[scale=0.5]{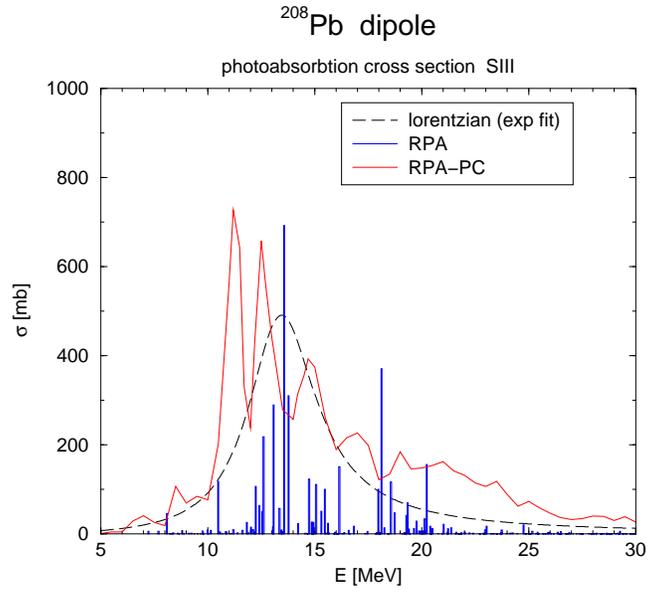}
\end{center}
\caption{Photoabsorption cross section for $^{208}$Pb.}
\label{fig1:pb}
\end{figure}

\begin{figure}[htbp]
\begin{center}
\includegraphics[scale=0.5]{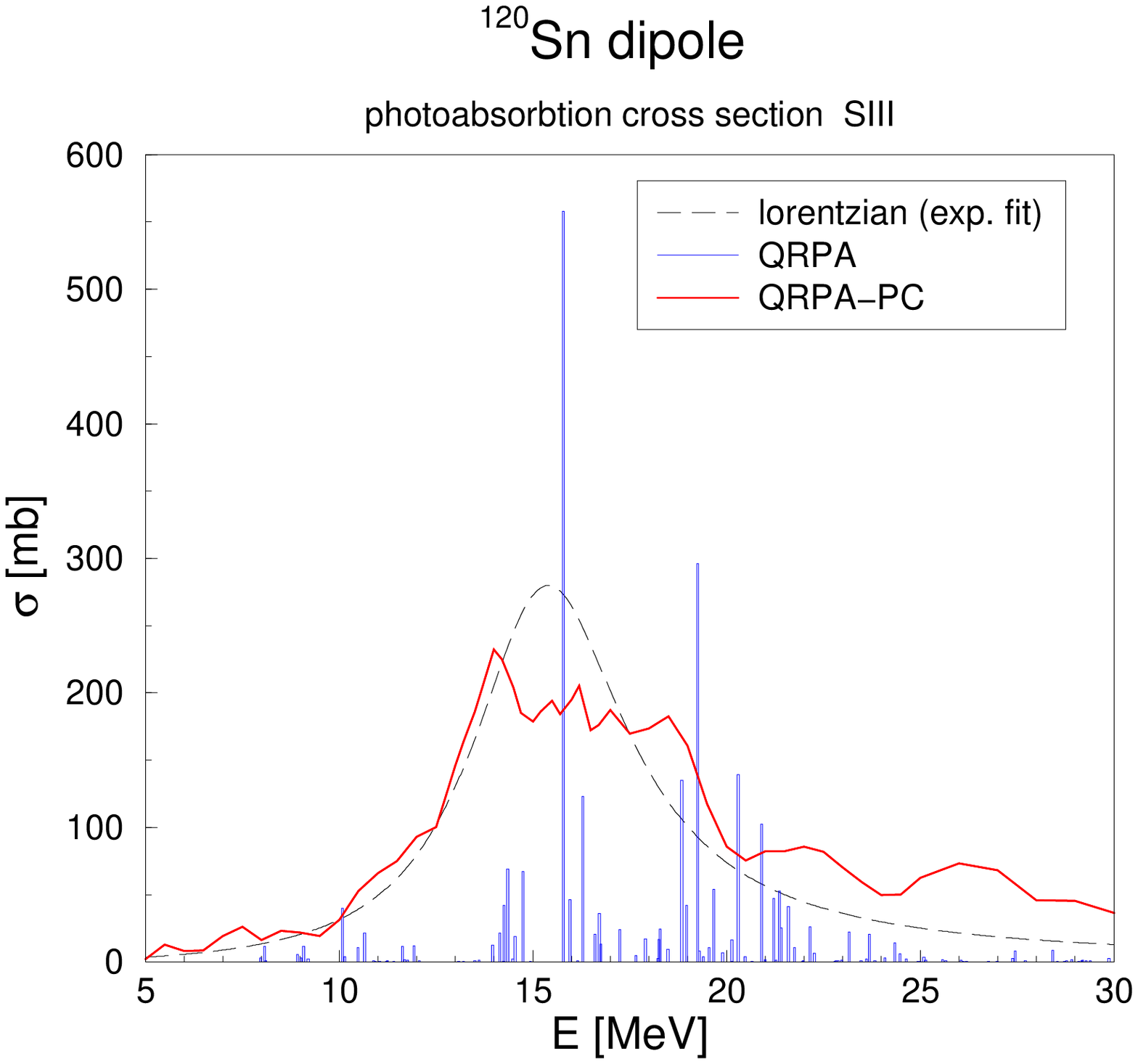}
\end{center}
\caption{Photoabsorption cross section for $^{120}$Sn.}
\label{fig2:120}
\end{figure}

\begin{figure}
\begin{center}
\includegraphics[scale=0.5]{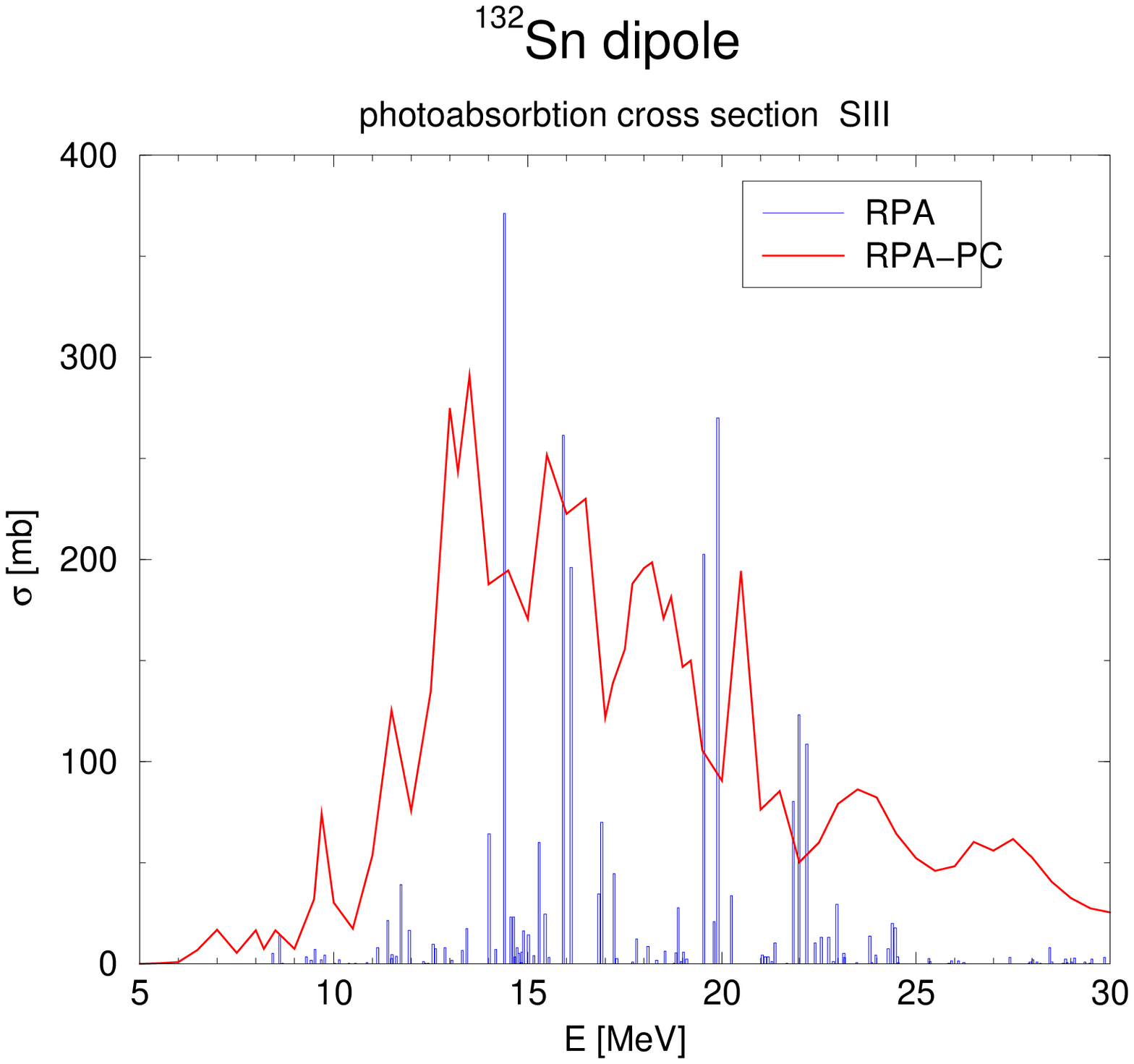}
\end{center}
\caption{Photoabsorption cross section for $^{132}$Sn.}
\label{fig3:132}
\end{figure}

\begin{figure}[htbp]
\begin{center}
\includegraphics[scale=0.5,angle=-90]{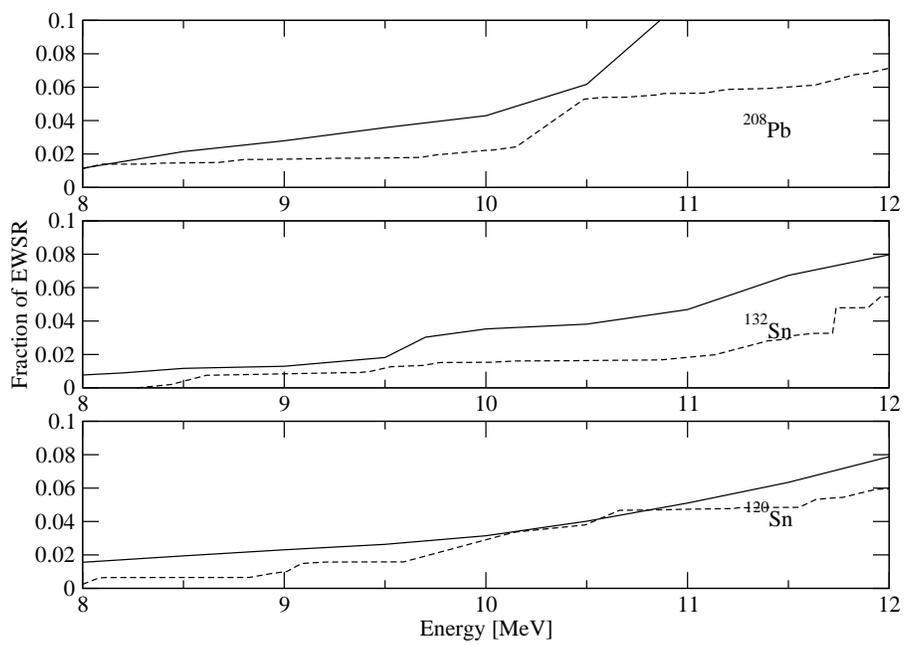}
\end{center}
\caption{Low-lying dipole strength in the three isotopes studied. The figure
has the same pattern of Fig. 1, but it refers only to the interval 8--12
MeV.}
\label{fig:ewsr2}
\end{figure}

\begin{figure}[htbp]
\begin{center}
\includegraphics[scale=1.0]{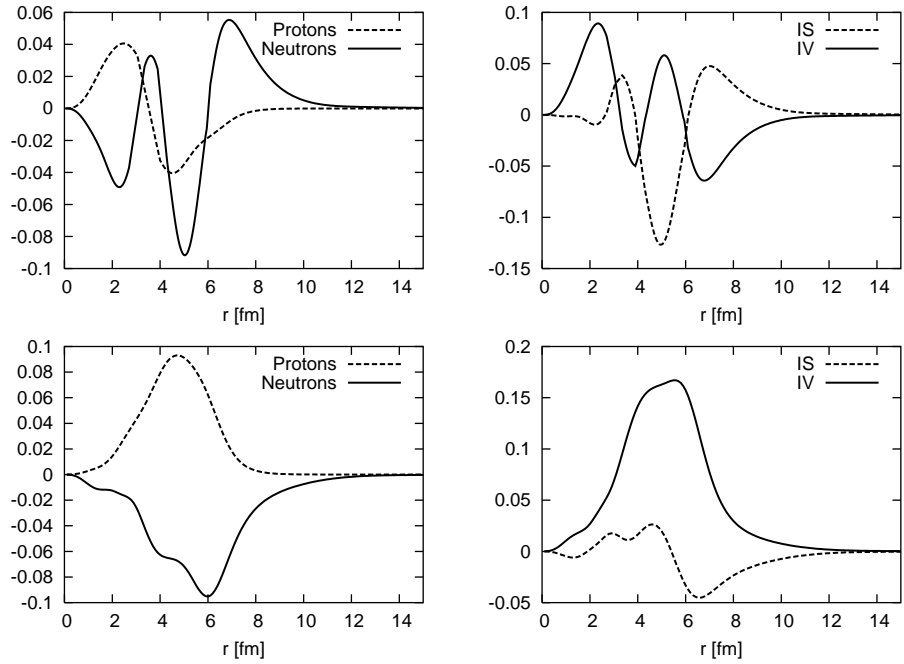}
\end{center}
\caption{Transition densities in $^{132}$Sn at the level of RPA-PC. 
The lower panels correspond
to the GDR at 13.5 MeV while the two upper panels to the low-lying state at
9.7 MeV MeV. Proton/neutron (isoscalar/isovector) transition densities are shown
in the left (right) side.}
\label{fig4:td}
\end{figure}

\end{document}